# On the Secure Degrees of Freedom in the $K$-User Gaussian Interference Channel

Onur Ozan Koyluoglu, Hesham El Gamal
Department of Electrical and Computer Engineering
The Ohio State University
Columbus, OH 43210, USA
Email: {koyluogo,helgamal}@ece.osu.edu

Lifeng Lai, H. Vincent Poor
Department of Electrical Engineering
Princeton University
Princeton, NJ 08544, USA
Email: {llai,poor}@princeton.edu

*Abstract*— This paper studies the $K$-user Gaussian interference channel with secrecy constraints. Two distinct network models, namely the interference channel with confidential messages and the one with an external eavesdropper, are analyzed. Using interference alignment along with secrecy pre-coding at each transmitter, it is shown that each user in the network can achieve non-zero secure Degrees of Freedoms (DoFs) in both scenarios. In particular, the proposed coding scheme achieves $\frac{K-2}{2K-2}$ secure DoFs for each user in the interference channel with confidential messages model, and $\frac{K-2}{2K}$ secure DoFs in the case of an external eavesdropper. The fundamental difference between the two scenarios stems from the lack of channel state information (CSI) about the external eavesdropper. Remarkably, the results establish the *positive impact* of interference on the secrecy capacity of wireless networks.

## I. INTRODUCTION

Recently, there has been a growing interest in the design and analysis of secure wireless networks based on information theoretic principles. For example, the secrecy capacity of networks involving relay nodes is studied in [1], [2], while the secrecy capacity of the wiretap channel with feedback is studied by [3]. On the other hand, multiple access channels with secrecy constraints were analyzed in [4], [5] and the broadcast channel model was investigated in [6]. Finally, the positive impact of multiple antennas and multi-path fading was established in [7], [8], [9], [10].

In this work, we consider the $K$-user Gaussian interference channel with secrecy constraints under a frequency-selective fading model. Without the secrecy constraints, it has been recently shown that a $\frac{1}{2}$ degrees of freedom (DoF) per orthogonal dimension is achievable for each source-destination pair in this network [11]. The achievability of this result was based on the *interference alignment* idea, by which the interfering signals are aligned to occupy a subspace orthogonal to the intended signals at each receiver. However, the impact of secrecy constraints on the degrees of freedom in this model has not yet been characterized. In fact, the only relevant work, which we are aware of, is the study of the two-user discrete memoryless interference channels with confidential messages [12], [13], [14] (our frequency selective model is fundamentally different from this discrete *memoryless* scenario).

In this work, we consider two distinct models. In the first scenario, one needs to ensure the *confidentiality* of each message from all non-intended receivers in the network. Towards this goal, we employ an interference alignment scheme along with secrecy pre-coding at each transmitter. The interference alignment scheme allows for aligning the non-intended signals, which are to be secured, at each receiver in the network. Then, the secrecy pre-coding ensures that the resulting multiple access channel (from the $K-1$ interfering users) does not reveal any information about each non-intended message. This way, we show that $\frac{1}{4}$ secure DoF per orthogonal dimension is achievable for each user in the three-user Gaussian interference channel with confidential messages. We then generalize our results to the $K$-user Gaussian interference channel showing that $\frac{K-2}{2K-2}$ secure DoFs are achievable for each user. In the second scenario, we study the external eavesdropper model in which the network users need only to secure their messages from the external eavesdropper. Here, the fundamental challenge is the lack of channel state information (CSI) about the external eavesdropper fading coefficients. Despite this fact, it is shown that $1/2-1/K$ DoFs is achievable in an ergodic setting. This result provides evidence on the diminishing gain resulting from knowing the instantaneous CSI of the eavesdropper *a-priori*. Interestingly, comparing our results with those obtained for point-to-point case [9], [10] reveals the positive impact of interference on the secrecy capacity of the network. The basic idea is that the coordination between several source-destination pairs allows for *hiding* the information in the background interference. Finally, we note that this paper only contains a sketch of one proof summarizing the main idea underlying our results. Interested readers can refer to the journal version of the paper for the detailed proofs.

## II. SYSTEM MODEL

*A. $K$-User Gaussian Interference Channel with Confidential Messages*

We consider a $K$-user Gaussian interference network, comprised of $K$ transmitter-receiver pairs, with confidential mes-

Hesham El Gamal also serves as the Director for the Wireless Intelligent Networks Center (WINC), Nile University, Cairo, Egypt.
This research was supported in part by the US NSF under Grants CCF-07-28762, CCF-03-46887, ANI-03-38807, CNS-06-25637, and CCF-07-28208.

sages. We represent the channel output at the $i$th receiver as follows:

$$Y_i(f,t) = \sum_{k=1}^{K} h_{ik}(f) X_k(f,t) + Z_i(f,t), \quad (1)$$

where $f \in \{1, \cdots, F\}$ denotes the frequency slot index, $t \in \{1, \cdots, n\}$ denotes the time slot index, $X_k(f,t)$ is the transmitted symbol of user $k$ at frequency slot $f$ during time $t$, and $Z_i(f,t) \sim \mathcal{N}(0,1)$ is the Gaussian noise at receiver $i$. We assume that the channel coefficients are randomly generated according to a continuous distribution and are fixed during the communication period. We also assume that the channel coefficients are known at every node in the network. [1]

Using the extended channel notation similar to [11], we write the received vector at receiver $i$ during time $t$ as follows:

$$\bar{\mathbf{Y}}_i(t) = \sum_{k=1}^{K} \mathbf{H}_{i,k} \bar{\mathbf{X}}_k(t) + \bar{\mathbf{Z}}_i(t), \quad (2)$$

where $\bar{\mathbf{Y}}_i(t) = [Y_i(1,t) \cdots Y_i(F,t)]^T$ is the $F \times 1$ column vector of received signal at user $i$ in time $t$, $\bar{\mathbf{Z}}_i(t) = [Z_i(1,t) \cdots Z_i(F,t)]^T$ is the $F \times 1$ column vector of receiver noise at user $i$, $\mathbf{H}_{i,k}$ is the $F \times F$ diagonal matrix of channel coefficients from transmitter $k$ to receiver $i$, and $\bar{\mathbf{X}}_k(t)$ is the $F \times 1$ column vector of transmitted symbols of user $k$, for some $i, k \in \mathcal{K}$.

We assume that each transmitter $k \in \mathcal{K}$ has a secret message $W_k$ which is to be secured from the remaining $K-1$ receivers while it can be decoded at the intended receiver $k$ with vanishingly small error probability. Considering $F$ frequency channels, which resembles parallel interference channels, our $(n, F, M_1, \cdots, M_K)$ secret codebook has the following components:

1) The secret message sets $\mathcal{W}_k = \{1, \cdots, M_k\}$ for transmitters $k \in \mathcal{K}$.

2) Encoding functions $f_k(.)$ at transmitters $k \in \mathcal{K}$, which map the secret messages to the transmitted symbols, $f_k : w_k \to (\bar{\mathbf{X}}_k(1), \cdots, \bar{\mathbf{X}}_k(n))$ for each $w_k \in \mathcal{W}_k$. At encoder $k$, each codeword is designed according to the transmitter's average long-term power constraint $P$, i.e.,

$$\frac{1}{nF} \sum_{f=1}^{F} \sum_{t=1}^{n} (X_k(t,f))^2 \leq P, \ k \in \mathcal{K}.$$

3) Decoding functions $\phi_k(.)$ at receivers $k \in \mathcal{K}$ which map the received symbols to estimates of the messages: $\phi_k(\mathbf{Y}_k) = \hat{W}_k, k \in \mathcal{K}$, where $\mathbf{Y}_k = \{\bar{\mathbf{Y}}_k(1), \cdots, \bar{\mathbf{Y}}_k(n)\}$.

Reliability of the transmission of user $k$ is measured by $P_{e,k}$, where

$$P_{e,k} = \frac{1}{M_k} \sum_{w_k \in \mathcal{W}_k} \Pr\{\phi_k(\mathbf{Y}_k) \neq w_k | w_k \text{ is sent}\}.$$

---
[1] We have the following notation in this work. Matrices are represented with bold capital letters ($\mathbf{X}$) and vectors are denoted as bold capital letters with bars or tildes (for example, $\bar{\mathbf{X}}$ and $\tilde{\mathbf{X}}$). We define $\mathcal{K} \triangleq \{1, \cdots, K\}$ and denote $\mathbf{X}_\mathcal{S} \triangleq \{\mathbf{X}_k | k \in \mathcal{S}\}$ for some $\mathcal{S} \subset \mathcal{K}$. Also, $o(\log(\rho))$ means that $\lim_{\rho \to \infty} \frac{o(\log(\rho))}{\log(\rho)} = 0$.

We use equivocation rate to measure the security level. More specially, for receiver $i$, the equivocation rates for each $\mathcal{S} \subset \mathcal{K} - i$ is defined as

$$\frac{1}{nF} H(\mathcal{W}_\mathcal{S} | \mathbf{Y}_i),$$

where $\mathcal{W}_\mathcal{S}$ is the set of secret messages of the users in the set $\mathcal{S}$.

We say that the rate tuple $(R_1, \cdots, R_K)$ is achievable if, for any given $\epsilon > 0$, there exists an $(n, F, M_1, \cdots, M_K)$ secret codebook such that,

$$\frac{1}{nF} \log_2 M_k \geq R_k - \epsilon, \ \forall k \in \mathcal{K},$$
$$\max\{P_{e,1}, \cdots, P_{e,K}\} \leq \epsilon, \quad (3)$$

and

$$\sum_{k \in \mathcal{S}} R_k - \frac{1}{nF} H(\mathcal{W}_\mathcal{S} | \mathbf{Y}_i) \leq \epsilon, \ \forall i \in \mathcal{K}, \forall \mathcal{S} \subset \mathcal{K} - i.$$

We also say that the secure degrees of freedom (per orthogonal frequency-time slot) tuple $(\eta_1, \cdots, \eta_K)$ is achievable, if the rate tuple $(R_1, \cdots, R_K)$ is achievable and

$$\eta_k = \lim_{\rho_k \to \infty} \frac{R_k}{\frac{1}{2} \log(\rho_k)}$$

for $k \in \mathcal{K}$, in which $\rho_k$ denotes the SNR at receiver $k$. As the receivers have unit-variance noises we will have $\rho_k = P$ in the sequel.

*B. K-User Gaussian Interference Channel with an External Eavesdropper*

In this model, we assume the existence of an external eavesdropper in the $K$-user interference network. The eavesdropper, through a Multiple Access Channel (MAC) from the legitimate users, has the following received signal

$$Y_e(f,t) = \sum_{k=1}^{K} h_{ek}(f) X_k(f,t) + Z_e(f,t), \quad (4)$$

at frequency slot $f$ and time slot $t$. The signal received at each receiver is the same as (1). We assume that $h_{ik}(f)$ is known at all the nodes in the network, including the eavesdropper. On the other hand, we consider two different scenarios regarding to $h_{ek}(f)$. In the first scenario, we assume that $h_{ek}(f)$ is known at all the nodes. In the second scenario, we assume that $h_{ek}(f)$ is only known at the eavesdropper. Hence, in this more interesting scenario, the sources and receivers do not have channel state information of the eavesdropper. The components of the secret codebook of each transmitter in the network can be represented as stated above. However, in this network model each transmitter must secure its own message *only* from the external eavesdropper. Accordingly, we modify the secrecy requirement by considering the equivocation rate seen by the eavesdropper. We denote the observation at the eavesdropper as $\mathbf{Y}_e = \{\bar{\mathbf{Y}}_e(1), \cdots, \bar{\mathbf{Y}}_e(n)\}$, in which $\bar{\mathbf{Y}}_e(t)$ is defined similarly as $\bar{\mathbf{Y}}_i(t)$ for $t = 1, \cdots, n$; and we represent the equivocation rate for a subset of users $\mathcal{S} \subset \mathcal{K}$ as

$$\frac{1}{nF} H\left(\mathcal{W}_\mathcal{S} | \mathbf{Y}_e\right).$$

We say that the rate tuple $(R_1, \cdots, R_K)$ is achievable, if for any given $\epsilon > 0$, there exits an $(n, F, M_1, \cdots, M_K)$ secret codebook such that

$$\begin{aligned} \frac{1}{nF} \log_2 M_k &\geq R_k - \epsilon, \forall k \in \mathcal{K}, \\ \max\{P_{e,1}, \cdots, P_{e,K}\} &\leq \epsilon, \end{aligned} \quad (5)$$

and

$$\sum_{k \in \mathcal{S}} R_k - \frac{1}{nF} H\left(\mathcal{W}_\mathcal{S} | \mathbf{Y}_e\right) \leq \epsilon, \forall \mathcal{S} \subset \mathcal{K}.$$

We also define the achievable secure DoF tuple for the users in this network model as was done in the previous section.

### III. THE $K$-USER GAUSSIAN INTERFERENCE CHANNEL WITH CONFIDENTIAL MESSAGES

Considering receiver $i \in \mathcal{K}$ of the interference network with confidential messages and denoting its received observation as $\mathbf{Y}_i$, the following lemma relates the equivocation rate for the full message-set with the secrecy notion given in (3).

*Lemma 1:* For a given $\epsilon > 0$, consider receiver $i \in \mathcal{K}$. If

$$\sum_{k \in \mathcal{K} - i} R_k - \frac{1}{nF} H\left(\mathcal{W}_{\mathcal{K}-i} | \mathbf{Y}_i\right) \leq \epsilon,$$

then

$$\sum_{k \in \mathcal{S}} R_k - \frac{1}{nF} H\left(\mathcal{W}_\mathcal{S} | \mathbf{Y}_i\right) \leq \epsilon, \forall \mathcal{S} \subset \mathcal{K} - i.$$

Due to this observation, it is sufficient to consider the closeness of the equivocation rate of the full message set, i.e., $\frac{1}{nF} H\left(\mathcal{W}_{\mathcal{K}-i} | \mathbf{Y}_i\right)$, to the sum of the message rates in the set $\mathcal{K} - i$ for each $i \in \mathcal{K}$ to satisfy the secrecy constraints given in (3).

Considering the three-user interference channel, we let $F = 2m+1$ for some $m \in \mathbb{N}$. This is the $(2m+1)$ symbol extension of the channel considered in [11]. We now employ interference alignment precoding using the matrices $\bar{\mathbf{V}}_k$ of [11], so that the transmitted signals are of the form $\bar{\mathbf{X}}_k(t) = \bar{\mathbf{V}}_k \tilde{\mathbf{X}}_k(t)$, where $\tilde{\mathbf{X}}_k(t)$ represents the $(m+1) \times 1$ vector of transmitted streams from user 1 and $m \times 1$ vectors of transmitted streams from users 2 and 3. Here, the beamforming vectors of user $k$, i.e., $\bar{\mathbf{V}}_k$, is constructed according to the channel gains of the network users and are used for the transmission of the independently coded streams at each user, i.e., $\tilde{\mathbf{X}}_k(t)$. Denoting the number of streams at each user as $m_k$ (the length of the vector $\tilde{\mathbf{X}}_k(t)$), these beamforming vectors ($\bar{\mathbf{V}}_k$s) can be constructed to satisfy two properties: 1) The non-intended signals seen by each receiver (interfering signals that are transmitted by the remaining users) are aligned within some received signal subspace of that receiver. More precisely, column spaces of matrices $\mathbf{H}_{i,k} \bar{\mathbf{V}}_k$ for $k \in \mathcal{K} - i$ lie in a subspace of dimension $F - m_i$ at the receiver $i$. 2) Desired streams of a receiver spans a subspace orthogonal to the one spanned by the interfering signals at that receiver. In other words, the columns of $\mathbf{H}_{i,i} \bar{\mathbf{V}}_i$ are independent and are orthogonal to that of $\mathbf{H}_{i,k} \bar{\mathbf{V}}_k$ for each user $k \in \mathcal{K} - i$. This way, $F$ dimensional received signal space at each receiver is used to create $m_i$ interference free dimensions, spanned by the desired streams. Also, with an employment of the interference alignment scheme, it is possible to align the non-intended signals at each receiver. Hence, there exist two orthogonal subspaces at each receiver, one consisting of intended streams and the other consisting of the interfering signals transmitted by the remaining users.

Now, considering receiver 1 as the eavesdropper for the messages of users 2 and 3, we have the following MAC seen by the eavesdropper. Transmitter 2 has $m$ streams $\tilde{\mathbf{X}}_2(t)$; transmitter 3 has $m$ streams $\tilde{\mathbf{X}}_3(t)$. Now, due to interference alignment, these streams span low dimensional received signal space of receiver 1, while they seen as independent streams to receivers 2 and 3, respectively. Here, receivers 2 and 3 can decode their desired streams through zero-forcing their received signals, i.e., by multiplying $\bar{\mathbf{Y}}_i(t)$ with $(\mathbf{H}_{i,i} \bar{\mathbf{V}}_i)^H$ to obtain the interference free desired streams. At this point, we design the transmitted streams $\tilde{\mathbf{X}}_2(t)$ and $\tilde{\mathbf{X}}_3(t)$ according to the MAC seen by the eavesdropping receiver 1 to satisfy the secrecy requirement. Following a similar analysis for each receiver, we satisfy the secrecy requirements of the network and have the following result.

*Theorem 2:* For the three-user Gaussian interference channel with confidential messages, a secure DoF of $\eta_k = \frac{1}{4}$ per frequency-time slot for each user $k \in \{1, 2, 3\}$ is achievable almost surely.

*Proof:* Sketch of the proof is given in Section VI.A. ∎

Considering the general $K$-user interference channel, and following an analysis similar to the one given in Section VI.A, we obtain the following result.

*Theorem 3:* For the $K$-user Gaussian interference channel with confidential messages, a secure DoF of $\eta_k = \frac{K-2}{2K-2}$ per frequency-time slot for each user $k \in \mathcal{K}$ is achievable almost surely.

### IV. THE $K$-USER GAUSSIAN INTERFERENCE CHANNEL WITH AN EXTERNAL EAVESDROPPER

In this section, we consider the external eavesdropper case. First, it is easy to see that when the eavesdropper CSI is available *a-priori* at the different transmitter and receivers, then our results in the previous section extend naturally. Intuitively, one can imagine the existence of a virtual transmitter associated with the external eavesdropper transforming our $K$-user network into another one with $K + 1$-users. This way, one can achieve a secure DoF of $\eta_k = \frac{(K+1)-2}{2(K+1)-2} = \frac{K-1}{2K}$ per frequency-time slot for each user using the scheme of the previous section. In particular, for a two-user network with an external eavesdropper, it is possible to achieve $\frac{1}{4}$ secure DoFs if the eavesdropper CSI is available at the transmitters. More formally, we have the following result.

*Corollary 4:* For the $K$-user Gaussian interference channel with an external eavesdropper, a secure DoF of $\eta_k = \frac{K-1}{2K}$ per frequency-time slot for each user $k \in \mathcal{K}$ is achievable almost surely (assuming the availability of the eavesdropper CSI).

More interestingly, it is still possible to achieve positive secure DoF per user even without the eavesdropper CSI in **an ergodic setting**. To illustrate the idea, let's consider the $K = 3$ case. Here, the users of the network has $\frac{3m+1}{2m+1}$ total DoF while the MAC seen by the eavesdropper can only have $\frac{2m+1}{2m+1}$ DoF from its observations. Hence, via the appropriate choice of secrecy codebooks, the $\frac{m}{2m+1}$ additional DoF can be *evenly* distributed among the network users *on the average*, allowing for a $\frac{1}{6}$ secure DoF per user without any requirement on the eavesdropper CSI. In the general case, we have the following result (the detailed proof will be reported in the journal paper).

*Proposition 5:* For the $K$-user Gaussian interference channel with an external eavesdropper, a secure DoF of $\eta_k = \frac{1}{2} - \frac{1}{K}$ per frequency-time slot for each user $k \in \mathcal{K}$ is achievable in the ergodic setting (in the absence of the eavesdropper CSI).

Remarkable, a point-to-point channel with an external eavesdropper was shown to have zero DoF [10]. In our case, as we add more user pairs to the network, each user-pair is able to achieve non-zero DoF for $K \geq 2$. This seemingly surprising result is due to interference alignment which allows the transmitters to pack their information in low dimensionality subspace (as seen by the eavesdropper), and hence, impairing the ability of the eavesdropper to distinguish any of the transmitted messages efficiently.

## V. CONCLUSIONS

In this work, we have obtained achievability results for the secure DoFs in the $K$-user Gaussian interference channel with frequency/time selectivity. By using the interference alignment scheme with secrecy pre-coding at each transmitter, we have shown that each user in the network can achieve non-zero secure DoFs under the confidential message and external eavesdropper models. The most interesting aspect of our results is, perhaps, the discovery of the role of interference in increasing the secrecy capacity of multi-user wireless networks.

## VI. APPENDIX

*A. Sketch of the Proof of Theorem 2*

Here $R_k$ is chosen such that

$$\lim_{\rho_k \to \infty} \frac{R_k}{\frac{1}{2} \log(\rho_k)} = \frac{1}{4},$$

so that $\eta_k = \frac{1}{4}$ is the achievable DoF at user $k = 1, 2, 3$ for sufficiently high $n$ and $F$.

We fix an $m \in \mathbb{N}$ and let $m_1 = m + 1$ and $m_k = m$, $\forall k \neq 1$ to analyze coding over $F = 2m + 1$ frequency slots. We generate, for each user $k$, $2^{nm_k \left( \frac{F}{m_k}(R_k + R_k^x) \right)}$ codewords each of length $nm_k$, where the entries are independent and identically distributed (i.i.d.) $\sim \mathcal{N}(0, \frac{P-\epsilon}{c_k})$. These codewords are then randomly partitioned into $M_k = 2^{nFR_k}$ message bins, each of consisting of $M_k^x = 2^{nFR_k^x}$ codewords. Hence, for user $k$, the secrecy codebook we will use is of dimension $M_k \times M_k^x$, and an entry of the codebook will be represented by $\hat{\mathbf{X}}_k(w_k, w_k^x)$, where the bin index $w_k \in M_k$ is called as the secrecy message and the codeword index $w_k^x \in M_k^x$ is called as the randomization message. We remark that the secure transmission rate per orthogonal time and frequency slot is $R_k$ with this scheme.

Now, to send a message $w_k$, the transmitter $k$ looks into the bin $w_k \in \mathcal{W}_k$ and randomly selects a codeword in this bin, denoted by the index $w_k^x$, according to uniform distribution. It thus obtains $\hat{\mathbf{X}}_k(w_k, w_k^x)$ of size $m_k \times n$. We further denote the elements of this matrix as $\hat{\mathbf{X}}_k(w_k, w_k^x) = [\tilde{\mathbf{X}}_k(1), \cdots, \tilde{\mathbf{X}}_k(n)]$, where each element is an $m_k \times 1$ vector. Here, for each symbol time $t \in \{1, \cdots, n\}$, transmitter $k$ employs the interference alignment scheme, and maps $\tilde{\mathbf{X}}_k(t)$ to $\bar{\mathbf{X}}_k(t)$ via $\bar{\mathbf{X}}_k(t) = \bar{\mathbf{V}}_k \tilde{\mathbf{X}}_k(t)$. At this point, we remark that $c_k$ is chosen to satisfy the power constraint for each user: $c_k = \frac{tr(\bar{\mathbf{V}}_k \bar{\mathbf{V}}_k^H)}{F}$. So that, the power of the transmitted signal satisfies the long term power constraint for each user. Here, we choose the interference alignment vectors $\bar{\mathbf{V}}_k$ as given in [11].

Now, to satisfy the achievability requirements for every eavesdropper $i \in \mathcal{K}$, we choose the secrecy and randomization rates as below.

$$\begin{aligned} R_k &= \frac{1}{F} \min_{i \in \mathcal{K}} \left\{ I(\tilde{\mathbf{X}}_i(t); \bar{\mathbf{Y}}_i(t)) \right\} \\ &\quad - \frac{1}{(K-1)F} \max_{i \in \mathcal{K}} \left\{ I(\tilde{\mathbf{X}}_{\mathcal{K}-i}(t); \bar{\mathbf{Y}}_i(t)) \right\} \quad (6) \\ R_k^x &= \frac{1}{F} \min_{i \in \mathcal{K}, \mathcal{S} \subset \mathcal{K}-i} \left\{ \frac{1}{|\mathcal{S}|} I(\tilde{\mathbf{X}}_{\mathcal{S}}(t); \bar{\mathbf{Y}}_i(t) | \tilde{\mathbf{X}}_{\mathcal{K}-\mathcal{S}-i}(t)) \right\} \end{aligned}$$

Here, we note that, in the high SNR regime, as the channel matrices are of full rank with probability one, we have

$$I(\tilde{\mathbf{X}}_i(t); \bar{\mathbf{Y}}_i(t)) = \frac{m_i}{2} \log(P) + o(\log(P)),$$

$$I(\tilde{\mathbf{X}}_{\mathcal{K}-i}(t); \bar{\mathbf{Y}}_i(t)) = \frac{F - m_i}{2} \log(P) + o(\log(P)),$$

and

$$I(\tilde{\mathbf{X}}_{\mathcal{S}}(t); \bar{\mathbf{Y}}_i(t) | \tilde{\mathbf{X}}_{\mathcal{K}-\mathcal{S}-i}(t)) = \frac{r}{2} \log(P) + o(\log(P)),$$

where $r = m$ or $r = m + 1$ depending on $i$ and $\mathcal{S}$.

Hence the equations above become

$$R_k = \frac{m-1}{8m+4} \log(P) + o(\log(P)),$$

and

$$R_k^x = \frac{m}{8m+4} \log(P) + o(\log(P)),$$

respectively as $\rho_k = P \to \infty$.

We note that, with the employment of the interference alignment scheme, it is possible to have $m_i$ interference free desired streams at each receiver $i$. At this point, as the above rates are inside the capacity region for each user, $R_i + R_i^x \leq \frac{1}{F} I(\tilde{\mathbf{X}}_i(t); \bar{\mathbf{Y}}_i(t))$, $\forall i \in \mathcal{K}$, each user can decode its own streams. Hence, for a given $\epsilon$ there exists $n(\epsilon)$ such that for $n > n(\epsilon)$, we have $\max\{P_{e,1}, \cdots, P_{e,K}\} \leq \epsilon$.

Now, it remains to be shown that we can satisfy the secrecy requirements of the network given in (3) in the high SNR regime. Here, denoting the observation of the eavesdropper as $\mathbf{Y}_i$, we write the following.

$$\begin{aligned} H(\mathcal{W}_{\mathcal{K}-i}|\mathbf{Y}_i) &= H(\mathcal{W}_{\mathcal{K}-i}, \mathbf{Y}_i) - H(\mathbf{Y}_i) \\ &= H(\mathcal{W}_{\mathcal{K}-i}, \mathcal{W}_{\mathcal{K}-i}^x, \mathbf{Y}_i) - H(\mathcal{W}_{\mathcal{K}-i}^x|\mathcal{W}_{\mathcal{K}-i}, \mathbf{Y}_i) - H(\mathbf{Y}_i) \\ &= H(\mathcal{W}_{\mathcal{K}-i}) + H(\mathcal{W}_{\mathcal{K}-i}^x|\mathcal{W}_{\mathcal{K}-i}) + H(\mathbf{Y}_i|\mathcal{W}_{\mathcal{K}-i}, \mathcal{W}_{\mathcal{K}-i}^x) \\ &\quad - H(\mathcal{W}_{\mathcal{K}-i}^x|\mathcal{W}_{\mathcal{K}-i}, \mathbf{Y}_i) - H(\mathbf{Y}_i) \\ &= H(\mathcal{W}_{\mathcal{K}-i}) + H(\mathcal{W}_{\mathcal{K}-i}^x) - I(\mathcal{W}_{\mathcal{K}-i}^x, \mathcal{W}_{\mathcal{K}-i}; \mathbf{Y}_i) \\ &\quad - H(\mathcal{W}_{\mathcal{K}-i}^x|\mathcal{W}_{\mathcal{K}-i}, \mathbf{Y}_i), \end{aligned} \quad (7)$$

where the last equality follows from the fact that $H(\mathcal{W}_{\mathcal{K}-i}^x|\mathcal{W}_{\mathcal{K}-i}) = H(\mathcal{W}_{\mathcal{K}-i}^x)$ as the randomization (i.e., codeword) indices are independent of the message (i.e., bin) indices.

We now bound each term of (7). Firstly, we have

$$I(\mathcal{W}_{\mathcal{K}-i}^x, \mathcal{W}_{\mathcal{K}-i}; \mathbf{Y}_i) \leq I(\tilde{\mathbf{X}}_{\mathcal{K}-i}(1), \cdots, \tilde{\mathbf{X}}_{\mathcal{K}-i}(n); \mathbf{Y}_i)$$

due to the Markov chain

$$\{\mathcal{W}_{\mathcal{K}-i}, \mathcal{W}_{\mathcal{K}-i}^x\} \to \{\tilde{\mathbf{X}}_{\mathcal{K}-i}(1), \cdots, \tilde{\mathbf{X}}_{\mathcal{K}-i}(n)\} \to \mathbf{Y}_i.$$

Combining with the fact that

$$I(\tilde{\mathbf{X}}_{\mathcal{K}-i}(1), \cdots, \tilde{\mathbf{X}}_{\mathcal{K}-i}(n); \mathbf{Y}_i) \leq nI(\tilde{\mathbf{X}}_{\mathcal{K}-i}(t); \bar{\mathbf{Y}}_i(t)) + n\delta_n^{(1)}$$

we have

$$I(\mathcal{W}_{\mathcal{K}-i}^x, \mathcal{W}_{\mathcal{K}-i}; \mathbf{Y}_i) \leq nI(\tilde{\mathbf{X}}_{\mathcal{K}-i}(t); \bar{\mathbf{Y}}_i(t)) + n\delta_n^{(1)}, \quad (8)$$

where $\delta_n^{(1)} \to 0$ as $n \to \infty$.

Secondly, we have

$$\begin{aligned} H(\mathcal{W}_{\mathcal{K}-i}^x) &= \log\left(\prod_{k \neq i} M_k^x\right) = nF \sum_{k \in \mathcal{K}-i} R_k^x \\ &= nI(\tilde{\mathbf{X}}_{\mathcal{K}-1}(t); \bar{\mathbf{Y}}_1(t)). \end{aligned} \quad (9)$$

Finally, we can bound

$$H(\mathcal{W}_{\mathcal{K}-i}^x|\mathcal{W}_{\mathcal{K}-i}, \mathbf{Y}_i) \leq n\delta_n^{(2)}, \quad (10)$$

where $\delta_n^{(2)} \to 0$ as $n \to \infty$. Roughly speaking, this is due to the fact that the eavesdropper can decode randomization messages given the secret messages $\mathcal{W}_{\mathcal{K}-i}$ and the observation $\mathbf{Y}_i$ with the Fano's inequality.

Plugging the bounds (8), (9), and (10) to (7) and dividing both sides by $nF$, we have

$$\frac{1}{nF}\left(H(\mathcal{W}_{\mathcal{K}-i}) - H(\mathcal{W}_{\mathcal{K}-i}|\mathbf{Y}_i)\right) \leq \frac{\delta_{P,i} + \delta_n^{(1)} + \delta_n^{(2)}}{F}, \quad (11)$$

where

$$\delta_{P,i} \triangleq I(\tilde{\mathbf{X}}_{\mathcal{K}-i}(t); \bar{\mathbf{Y}}_i(t)) - I(\tilde{\mathbf{X}}_{\mathcal{K}-1}(t); \bar{\mathbf{Y}}_1(t)) \geq 0, \quad (12)$$

for $i = 1, 2, 3$.

Here, as $n \to \infty$ and $m \to \infty$, we can make $\frac{\delta_{P,i} + \delta_n^{(1)} + \delta_n^{(2)}}{F}$ arbitrarily small. Then, utilizing Lemma 1, we have readily satisfied

$$\sum_{k \in \mathcal{S}} R_k - \frac{1}{nF} H(\mathcal{W}_{\mathcal{S}}|\mathbf{Y}_i) \leq \epsilon, \; \forall i \in \mathcal{K}, \forall \mathcal{S} \subset \mathcal{K} - i. \quad (13)$$

To summarize, given an $\epsilon > 0$, there exist $n_0$ and $m_0$ such that, for any $n > n_0$ and $m > m_0$ we satisfy the achievability requirements given in (3) for a given $P$. As $P \to \infty$, by taking $m \to \infty$ appropriately, we have $R_k = \frac{1}{8}\log(P) + o(\log(P))$ for $k = 1, 2, 3$. Hence, each user in the three-user interference network is able to achieve $\frac{1}{4}$ secure DoF per orthogonal frequency and time dimension.